\documentclass[12pt]{article}
\usepackage{amssymb}
\usepackage{graphicx,epsfig}
\begin{document}

\title{\bf Soft ridge in proton-proton collisions}

\author{M.Yu. Azarkin, I.M. Dremin, A.V. Leonidov \\ {\small P.N. Lebedev Physics Institute, Moscow, Russia}}
\date{}
\maketitle

\begin{abstract}
It is shown that the soft mechanism of multiparticle production by Lund hadronic strings formed by colliding constituent degrees of freedom
generates a shape of angular correlations similar to the ridge structure  observed in the pp collisions at 7 TeV at the LHC.
\end{abstract}

\newpage

The recent observation of the ridge-like structure of the two-particle near side angular
correlations in high multiplicity events in $pp$
collisions at LHC at $\sqrt{s}=7 \; {\rm TeV}$ \cite{cms} provided, for the first time,
a possible evidence for the coherent nature of
multiparticle production in hadron collisions at high energies. The observed shape of
the correlation function $C(\Delta \eta,\Delta \phi)$ is
characterized by long-range correlations in $\Delta \eta$ spanning several units of
(pseudo)rapidity and well-localized azimuthal structure in
$\Delta \phi$ centered around $\Delta \phi=0$. Of particular interest is the possible
 connection of this phenomenon to the similar structure of
the long-range near side correlations earlier discovered in heavy ion collisions at RHIC
at $\sqrt{s}=200 \; {\rm GeV}$ \cite{star, phen, phob}.
Still earlier, before RHIC, experimental analysis of $C(\Delta \eta,\Delta \phi)$
did not reveal such ridge-like structures (e.g., see
\cite{egg,ajin} and the review paper \cite{ddk}).

Let us stress that the observation region of the CMS experiment favors production of comparatively soft particles (see also \cite{hwa}). The ridge
is filled in mostly by particles with transverse momenta in the range $1<p_T<3$ GeV/c which are emitted at comparatively large polar angles $\vert
\eta \vert <2.4$. It is easily derived that the longitudinal momenta are less than 17 GeV and the share of the initial momentum $x$ is less than
5$\cdot 10^{-3}$. If treated in terms of multiperipheral ladders (or strings) this implies that mostly particles from their central parts are in
charge of the ridge. The matrix elements are larger where the denominators of the propagators are smaller. It was shown in \cite{dkim} that these
denominators contain contributions proportional to $1-\cos \Delta \phi$ which vanish at $\Delta \phi =0$ and form a parabolically shaped maximum.
This favours coplanarity in $\Delta \phi$ noticed in some cosmic ray studies \cite{slav} as well.

Both the similarity of multiparticle production in hadron and nuclear collisions at high energies and the long-range correlations in rapidity are
quite natural in the framework of the quantitative approach to QCD at small $x$, that of the Color Glass Condensate, see, e.g., the recent reviews
\cite{gijv10,m10,dl10}. This point of view is supported by the new experimental evidence from the LHC showing that the inclusive density in
pseudorapidity in pp collisions at 7 TeV becomes comparable to that at RHIC while the size of the pp-interaction region at very high multiplicity
becomes also comparable to that in Au-Au as derived from Bose-Einstein correlations. Calculations of the angular correlation based on the
inclusive cross-section of two gluon production within the CGC paradigm were discussed, in two different Lorentz frames, in \cite{dum,KL10}. The
main conclusion of both papers is that the ridge structure in high-energy proton-proton collisions is a natural consequence of the high gluon
density characteristic for the wave functions of fast incoming protons. The quantitative comparison with the CMS data \cite{cms} was, however, not
presented. The corresponding description for the ridge in heavy ion collisions in terms of glasma flux tubes was given in
\cite{DGMV08,LM10,DGLV10}

In all of the above-mentioned descriptions of the ridge one considers boost invariant dynamics as a source of long-range rapidity correlations.
This part of the argument can be put on the solid basis. As to the localized structure in the azimuthal angular distance, the existing arguments
are still more qualitative than quantitative.

Returning again to the long-range correlations in rapidity let us note that multiperipheral graphs with reggeon quantum number exchange lead also
to rather small separation in pseudorapidities $\Delta \eta$ because of low transferred momenta. Larger separation of $\Delta \eta$ is expected if
the pomeron is exchanged at some steps in the ladder. Particles from such diffractive processes would be in charge of ridge regions at larger
$\Delta \eta$ thus elongating the ridge. In the related context, as also mentioned in \cite{T10}, the boost-invariant picture of particle
production and, therefore, long-range rapidity correlations, are also characteristic of the usual soft multiparticle production in the Lund string
model \cite{LUND} on which, in particular, a description of soft multiparticle production in the PYTHIA Monte-Carlo generator \cite{PYTHIA} is
based. Therefore it is natural to ask, whether the ridge can be generated by usual hadronic string's breaking as such. This question can be
answered by switching off all non-soft mechanisms of particle production possible in the version PYTHIA 6.4\footnote{More concretely, the
corresponding switch is ensured by setting the parameter MSTP(82) equal to zero.} In this setting the only sources of multiparticle production are
two hadronic strings stretching between the corresponding valence degrees of freedom.

The resulting correlation function generated with PYTHIA 6.4 for $\sqrt{s}=7\;{\rm TeV}$ with the cuts on pseudorapidity $2.0<|\eta|<4.0$, trigger
transverse momentum $p^{\rm trig}_\perp=1.0-3.0 \; {\rm GeV}$, away-side transverse momentum $p^{\rm as}_\perp=0.6-1.5 \; {\rm GeV}$ and
multiplicity $n_{\rm part} > 12$ is plotted, together with the corresponding experimental data from \cite{cms}, in Fig.~1\footnote{For convenience
we normalize the experimental data on the theoretical value at the away-side peak $\Delta \phi = \pi$. }. Let us stress that the cut $n_{\rm part}
> 12$ is in fact the {\it high multiplicity} cut for the multiparticle production having its origin only in the standard fragmentation of hadronic
strings with all hard mechanisms switched off whereas the experimentally observed ridge structure corresponds to the cut $n_{\rm part} > 110$
\cite{cms}. The possible mechanisms of amplification of string-like contribution making it potentially relevant for describing the experimental
data are discussed at the end of this note.
\begin{figure}[ht]
\centering
\includegraphics[height=0.4\textheight,width=0.8\textwidth]{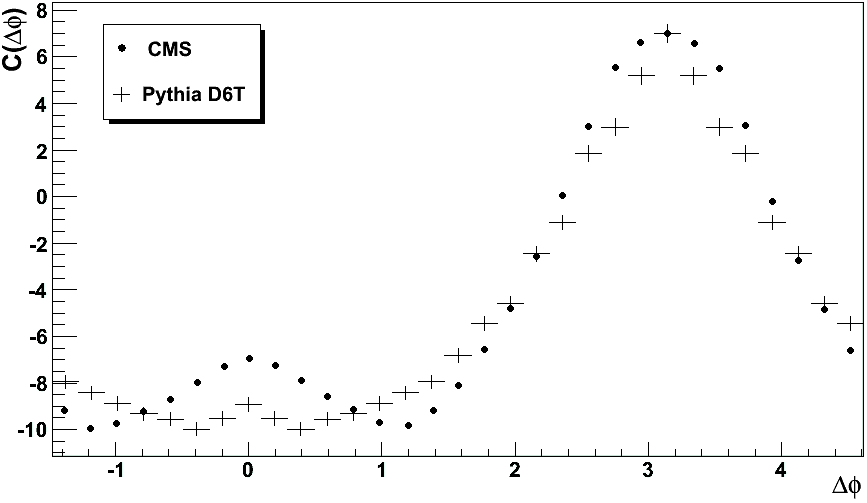}
\caption{ Comparison of the correlation functions for experimental data and MC: experimental data(dots), MC(crosses). MC data were obtained in the
conditions: $2.0<|\eta|<4.0$,  $p^{\rm trig}_\perp=1.0-3.0 \; {\rm GeV}$, $p^{\rm as}_\perp=0.6-1.5 \; {\rm GeV}$, $n_{\rm part} > 12$, 5.5M
events; data: $2.0<|\eta|<4.8$,  $p^{\rm trig}_\perp=1.0-3.0 \; {\rm GeV}$, $p^{\rm as}_\perp=1.0-3.0 \; {\rm GeV}$, $n_{\rm part} > 110$, 354k
events. Experimental data are normalized to MC simulation at the away-side peak $\Delta \phi = \pi$.}
\end{figure}

The result of the computation, shown in Fig.~1, clearly demonstrates that the standard multiparticle production through the Lund hadronic string
breaking does indeed generate a ridge-like structure which in its shape is quite similar to that observed in the experiment \cite{cms}. As
previously mentioned, the source of the long-range correlations in pseudorapidity in the model under consideration is in the (approximate) boost
invariance of the string decay. As to the emergence of the localized structure at $\Delta \phi = 0$ seen in Fig.~1, it is natural to assume that
it is related, on the event-by-event basis, to the existence of the special direction in the impact parameter plane given by the vector of impact
parameter of the colliding valence degrees of freedom for each string.

The result shown in Fig.~1 raises two interrelated questions. First, why one
does not see the ridge structure in the full version of PYTHIA
\cite{cms}? Second, what is the relation of the soft ridge seen in Fig.~1
to the ridge structure observed in the CMS experiment?

As of now we do not have definitive answers to these questions. At the qualitative level, the answer to the first questions is that the semi-hard
partonic showers, the second principal mechanism of multiparticle production in PYTHIA, wash out the fragile soft ridge structure generated by the
strings corresponding to the scattering of the valence degrees of freedom. As to the second question, the main issue is whether one can rescale
the soft ridge phenomenon that is naturally characterized by relatively low multiplicities to the multiplicity regime in which the ridge is
observed at the LHC. In our opinion the possible link could be provided by the important role presumably played by glasma flux tubes (or, in a
related context, large longitudinal fields within the color ropes \cite{colrope,TGBGW10}) in multiparticle production at high energies. The
relevant configurations are very similar to the Lund strings, the main distinction being in the strength of the chromoelectric fields within the
corresponding flux tubes and, therefore, enhanced particle production.

This work was supported by RFBR grants 09-02-00741 and by the RAN-CERN program.


\begin{thebibliography}{99}

\bibitem{cms}
V. Khachatryan et al. [CMS Collaboration], JHEP {\bf 1009}, 091 (2010).
\bibitem{star}
B.I. Abelev et al. [STAR Collaboration], Phys. Rev. C {\bf 80}, 064912 (2009).
\bibitem{phen}
A. Adare et al. [PHENIX Collaboration], Phys. Rev. Lett. {\bf 104}, 252301
(2010).
\bibitem{phob}
B. Alver et al. [PHOBOS Collaboration], Phys. Rev. Lett. {\bf 104}, 142301
(2010).
\bibitem{egg}
K. Eggert et al., Nucl. Phys. B {\bf 86}, 201 (1975).
\bibitem{ajin}
I. Ajinenko et al. [NA22 Collaboration], Z. Phys. C {\bf 58}, 357 (1993).
\bibitem{ddk}
E.A. DeWolf, I.M. Dremin, and W. Kittel, Phys. Rep. {\bf 270} 1 (1996).
\bibitem{hwa}
R.C. Hwa and C.B. Yang, arXiv:1011:0965 [hep-ph].
\bibitem{dkim}
I.M. Dremin and V.T. Kim, Pis'ma v ZhETF {\bf 92} 720 (2010).
\bibitem{slav}
S.A. Slavatinsky, Nucl. Phys. B Suppl. {\bf 122}, 3 (2003).
\bibitem{gijv10}
F. Gelis, E. Iancu, J. Jalilian-Marian, R. Venugopalan, arXiv:1002.0333 [hep-ph]
\bibitem{m10}
L. McLerran, arXiv:1011.3203 [hep-ph]
\bibitem{dl10}
I.M. Dremin and A.V. Leonidov, Physics-Uspekhi {\bf 180}, 1167 (2010); arXiv:1006.4607 [hep-ph]
\bibitem{dum}
A. Dumitru, K. Dusling, F. Gelis, J. Jalilian-Marian, T. Lappi, R. Venugopalan, arXiv:1009.5925 [hep-ph].
\bibitem{KL10}
A. Kovner, M. Lublinsky, arXiv:1012.3398 [hep-ph]
\bibitem{DGMV08}
A. Dumitru, F. Gelis, L. McLerran, R. Venugopalan, Nucl. Phys, A {\bf 810} (2008), 91
\bibitem{LM10}
T. Lappi, L. McLerran, Nucl. Phys. A {\bf 832} (2010), 330
\bibitem{DGLV10}
K. Dusling, F. Gelis, T. Lappi, R. Venugopalan, Nucl. Phys, A {\bf 836} (2010), 159
\bibitem{T10}
T.A. Trainor, arXiv:1012.2371 [hep-ph]
\bibitem{LUND}
B. Andersson, G. Gustaffson, G. Ingelman, T. Sjostrand, Phys. Rept. {\bf 97} (1983), 31
\bibitem{PYTHIA}
T. Sjostrand, S. Mrenna, P. Skands  {\it JHEP}\ 0605:026 (2006)
\bibitem{colrope}
T.S. Biro, H.B. Nielsen, J. Knoll, Nucl. Phys. B {\bf 245} (1984), 449
\bibitem{TGBGW10}
V. Topor Pop, M. Gyulassy, J. Barette, C. Gale, A. Waburton, arXiv:1010.5439 [hep-ph]





\end{thebibliography}
\end{document}